# MITIGATING DEEP LEARNING NETWORKS VULNERABILITIES FROM ADVERSARIAL EXAMPLES ATTACK IN THE CYBERSECURITY DOMAIN

Chris Einar Mirasol San Agustin
Institute of Electrical and Electronic Engineers
sanagustin@ieee.org

**ABSTRACT**

Deep learning models are known to solve classification and regression problems by employing a number of epoch and training samples on a large dataset with optimal accuracy. However, that doesn't mean they are attack-proof or unexposed to vulnerabilities. Newly deployed systems particularly on a public environment (i.e public networks) are vulnerable to attacks from various entities. Moreover, published research on deep learning systems (Goodfellow et al., 2014) have determined a significant number of attacks points and a wide array of attack surface that has evidence of exploitation from adversarial examples. Successful exploit on these systems could lead to critical real world repercussions. For instance, (1) an adversarial attack on a self-driving car running a deep reinforcement learning system yields a direct misclassification on humans causing untoward accidents.(2) a self-driving vehicle misreading a red light signal may cause the car to crash to another car (3) misclassification of a pedestrian lane as an intersection lane that could lead to car crashes. This is just the tip of the iceberg, computer vision deployment are not entirely focused on self-driving cars but on many other areas as well—that would have definitive impact on the real-world. These vulnerabilities must be mitigated at an early stage of development. It is imperative to develop and implement baseline security standards at a global level prior to real-world deployment.

Deep learning algorithms have seen their deployment in multiple industries in an upward trajectory and will continue to increase in the upcoming years due to the (1) enhancements on learning algorithms and tools, (2) improved research by skilled engineers and scientists, (3) computing power. The rise of Deep Learning implementation will result to a larger attack surface for adversarial attacks. These scenarios embodies the vulnerabilities not just on deep learning but on a larger scale—the entire AI ecosystem. This paper will demonstrate the methodologies on mitigating adversarial attacks on deep learning systems—enhancing robustness to the system.

# 1 INTRODUCTION

Adversarial examples have immensely fooled classifiers into misclassification of training dataset—adding random noise to the input data,single step and multi-step attacks—have evidently made their way in compromising deep learning systems. Moreover, targeting training examples by perturbing the image identity (i.e,. car is misclassified as a dog) are some of the widely published techniques in attacking the system. Despite it's high accuracy, these systems are vulnerable through a wide range of attack surface that is proven to be exploitable. In this paper, we aim to develop an intuition on mitigating adversarial examples to further enhance the robustness of deep learning systems. A networked deep learning system contains a number of entry and exit points to and from the network and must be mitigated from the early stages of development or prior to the deployment. Empirically, we have found that these entry points increases the probability of a successful adversarial attack. We propose a vulnerability rating score (probability of a successful exploit) for each vulnerability found on a deep learning system and set a global standard on mitigating each vulnerability. Furthermore, It is important to understand the ambiguous entry points and where mitigation must be in place to reduce the risk of adversarial examples. Figure 1 shows the base score metrics each CVE, we aim to standardized all Machine Learning vulnerabilities and set a base score metric for each vuln.

*Nvd.nist.gov*

Intuitively, a networked system must have in depth security defenses on each attack surface—from the physical layer to the application layer. We also propose that deep learning deployment must be designed and implemented aligned with the IEEE security and networking standards to reduce the risk of exploitation.

## 1.1 WHITE BOX AND BLACK BOX TECHNIQUES

There are two types of adversarial attacks: white box and blackbox techniques. White box based attacks leverages an attacker's knowledge of the entire network wherein the attacker has gainful insight of the network's architecture (i.e input data, training examples hyperparameters and number of layers). Whereas, black box attacks are performed wherein an attacker has only partial knowledge of the ML system architecture. Concretely, an adversarial attack on a self-driving car and facial recognition software on the real world can manifest through: (1) A system misclassifying a human face by using an infrared adversarial invisible mask that tricks the system and leads to unrecognizable face detection (Zhe Zhou et al., 2018). (2) misclassification of street signs by printing out adversarially constructed image. (3) Targeted classification altered image identity wherein the target image is incorrectly identified. These probable occurrences have real world implications and can be detrimental to the massive deployment of deep learning systems in production environment and may even harm humans to a greater extent.

## 1.2 NETWORKED DEEP LEARNING SYSTEMS

Intuitively, designers have to take into consideration the variations of deep learning implementation: for one a self-driving car—is most likely to be networked on a public environment (i.e,. The internet). This opens up a ton of opportunities for attackers that increases the probability of a successful exploit not just from adversarial examples but from different forms of malware. Since the code resides on the application layer and hosted locally within the car itself, there has to be security measures in place on the physical layer which is inside the car. Some important question needs to be answered (1) Does perimeter security (i.e,. Firewall) has empirical value on deployment? (2) How does endpoint security fit into the equation (i.e,. Antivirus) to defend against endpoint attack. (3) There have been some known security loopholes on 5G networks (Jover et al., 2019) which will power self-driving cars internet connection. How does this impact the robustness of deep networks? It is quite evidenced that deep neural networks will not only be vulnerable to adversarial examples but also to a greater extent they are geared towards the cybersecurity and network space that allows them to be vulnerable to any attack just like a software application which is what they are.

**2 ADVERSARIAL TRAINING DEFENSE**

Adversarial training is one of the known methods in mitigating adversarial examples—making the network more robust from white hat and black hat attacks and is highly considered as the most effective way of mitigation (Kurakin et al., 2018). In adversarial training the network is being trained to classify images with clean examples and perturbed or adversarial examples—allowing a baseline for error classification.

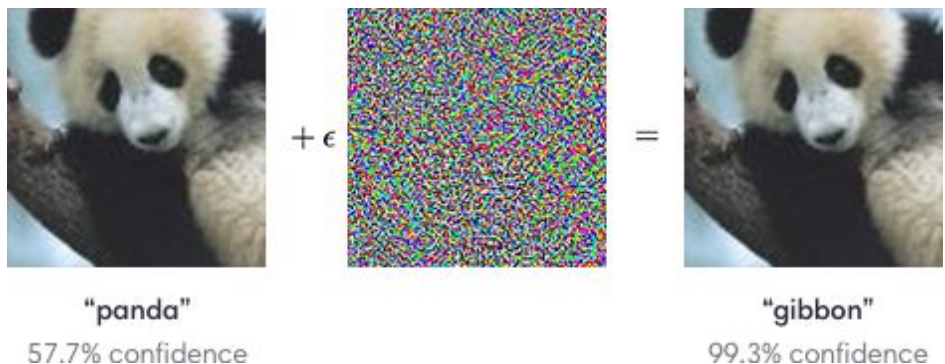

From the image shown above, a panda on the left and a gibbon on the right. An attacker can add a random noise or a minor perturbation that can result into tricking the classifier of categorizing the panda as a gibbon. In Adversarial training, we configure the classifier as having the best and worst case scenario of image classification—we train the model into classifying the image as its best case: a panda and a worst case: a gibbon. This is an important aspect of training, in which any perturbation can be deflected by adversarial training—allowing a more robust system that can resist adversarial examples. Furthermore, it has been known that adversarially trained models exploited by single step attacks to generate adversarial examples are easier to classify as well as for undefended model (Goodfellow et al., 2018). Definitively, adversarial training not only learns to deflect the attack but also make the attack performs at a worse level (Goodfellow et al., 2018).

## 2.1 GENERATING ADVERSARIAL EXAMPLES

Adversarial example generation can be categorized into (1) single step attack where there is only a single gradient computation and (2) multi step or iterative attack where there are multiple gradient computations iteration. The objective of every adversarial example is to have a high error rate on the loss function $L(X_n + r_n, y_{true\ n}; \theta)$ for each image $X_n$ (Zhou Ren et al., 2018), taking into account that the image generated is similar to the image from the training example. Furthemore, adding randomization layers on the model's architecture has been found to be successful defense in adversarial examples particularly in multi-step iterative attacks, in stark contrast to adversarial training where it has evidence of having a high success rate in defending single step attacks (Goodfellow et al., 2018)

It is important to note that these attacks can be exploited using white box and black box techniques and evidently—security against white-box attacks is the main goal because of the attackers access and knowledge of the system, although black-box security has more emphasis in developing a baseline goal for deployed ML models.

Below we list the methods in generating adversarial examples and its impact on the network.

**3 Fast Gradient Sign Method (FGSM)**. (Goodfellow et al., 2014b)

FGSM is considered as a single step attack where a single gradient is computed to generate the adversarial example. FGSM leverages the following formula:

$x\ adv\ FGSM := x + \varepsilon \cdot sign\left(\nabla xL(h(x), y_{true})\right).$

$$x^{adv} = x + \varepsilon \cdot \text{sign}(\nabla_x J(x, y_{true})),$$

where

$x$ is the input (clean) image,

$x^{adv}$ is the perturbed adversarial image,

$J$ is the classification loss function,

$y_{true}$ is true label for the input $x$.

FSGM in white hat based attacks targets perturbation on input data therefore resulting into a higher loss based on the same backpropagated gradients. It is architected to attack deep learning networks by the way the networks learn-gradients. Intuitively, there has been some notion that in a black-box setting where an attacker does not have full access to the model's architecture. A transferrable attack can be propagated from a trained adversarial network that could be transferred to the targeted network

## 4 TRANSFER ADVERSARIAL ATTACK

There has been formal and empirical evidence that adversarial example can transfer to more than one model (Papernot et al., 2017). In a black box setting (where an attacker does not have full access of the model's architecture) an attacker can train a surrogate model that has the same input training examples as the targeted model. This leads to a higher probability of successfully exploiting the target model using a surrogate model. Furthermore, input data generated from one model performing the same task can be transferred to another model. Transferring an attack has limitations (Boneh et al., 2017). Concretely, it has been proven that transferability of model-agnostic perturbations. As we can see from the image below, a small perturbation on an input image causes a direct misclassification on the training example.

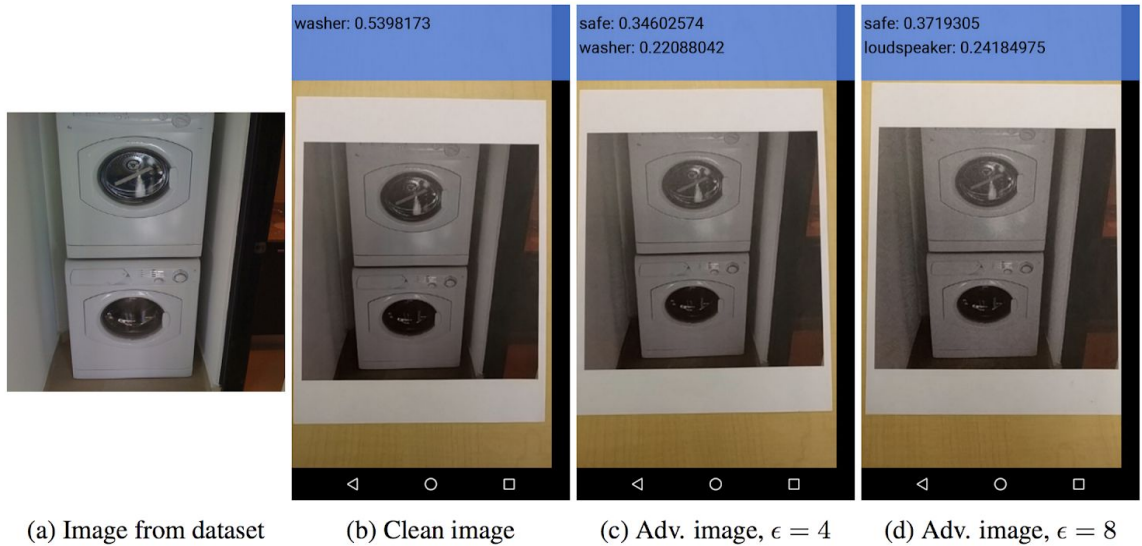

(a) Image from dataset  (b) Clean image  (c) Adv. image, $\epsilon = 4$  (d) Adv. image, $\epsilon = 8$

**5 Single-Step Least-Likely Class Method (Step-LL).** This variant of FGSM introduced by Kurakin et al. (2017a;b) targets the least-likely class, , yLL = arg min{h(x)}:

$$x^{adv}_{LL} := x - \epsilon \cdot \text{sign}(\nabla_x L(h(x), y_{LL})).$$

**6 Iterative Attack (I-FGSM or Iter-LL).** This method iteratively applies the FGSM or Step−LL $k$ times with step-size $\alpha \geq \epsilon/k$ and projects each step onto the $\ell_\infty$ ball of norm $\epsilon$ around x. It uses projected gradient descent to solve the maximization in (1). For fixed $\epsilon$, iterative attacks induce higher error rates than single-step attacks, but transfer at lower rates (Kurakin et al., 2017a;b).

**7 DeepFool:** DeepFool (Moosavi-Dezfooli et al., 2016) is an iterative attack method which finds the minimal perturbation to cross the decision boundary based on the linearization of the classifier at each iteration. Any lp-norm can be used with DeepFool, and we choose l2-norm for the study in this paper.

**8 Carlini & Wagner (C&W):** C&W (Carlini & Wagner, 2017) is a stronger iterative attack method proposed recently. It finds the adversarial perturbation $r_n$ by using an auxiliary variable $\omega_n$ as $r_n = \frac{1}{2}(\tanh(\omega_n + 1)) - X_n$. (2) Then the loss function optimizes the auxiliary variable $\omega_n$ $\min_{\omega_n} ||\frac{1}{2}(\tanh(\omega_n) + 1) - X_n|| + c \cdot f(\frac{1}{2}$

$(\tanh(\omega_n) + 1))$. (3) The function $f(\cdot)$ is defined as $f(x) = \max(Z(x)_{y_{true}} - \max\{Z(x)_i : i \neq y_{true}\}, -k)$, (4) where $Z(x)_i$ is the logits output for class $i$, and $k$ controls the confidence

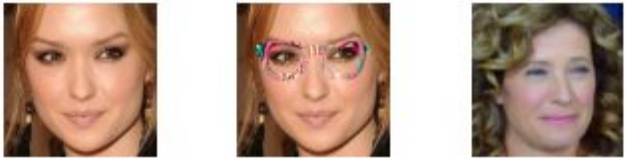

Figure 5: An example of adversarial eyeglass frame against Face Recognition System [67] Kos et al. extended.

Test results from (Goodfellow et al., 2017).

|  |  | Clean | $\epsilon = 2$ | $\epsilon = 4$ | $\epsilon = 8$ | $\epsilon = 16$ |
|---|---|---|---|---|---|---|
| Baseline (standard training) | top 1 | 78.4% | 30.8% | 27.2% | 27.2% | 29.5% |
|  | top 5 | 94.0% | 60.0% | 55.6% | 55.1% | 57.2% |
| Adv. training | top 1 | 77.6% | 73.5% | 74.0% | 74.5% | 73.9% |
|  | top 5 | 93.8% | 91.7% | 91.9% | 92.0% | 91.4% |
| Deeper model (standard training) | top 1 | 78.7% | 33.5% | 30.0% | 30.0% | 31.6% |
|  | top 5 | 94.4% | 63.3% | 58.9% | 58.1% | 59.5% |
| Deeper model (Adv. training) | top 1 | 78.1% | 75.4% | 75.7% | 75.6% | 74.4% |
|  | top 5 | 94.1% | 92.6% | 92.7% | 92.5% | 91.6% |

| Adv. method | Training |  | Clean | $\epsilon = 2$ | $\epsilon = 4$ | $\epsilon = 8$ | $\epsilon = 16$ |
|---|---|---|---|---|---|---|---|
| Iter. l.l. | Adv. training | top 1 | 77.4% | 29.1% | 7.5% | 3.0% | 1.5% |
|  |  | top 5 | 93.9% | 56.9% | 21.3% | 9.4% | 5.5% |
|  | Baseline | top 1 | 78.3% | 23.3% | 5.5% | 1.8% | 0.7% |
|  |  | top 5 | 94.1% | 49.3% | 18.8% | 7.8% | 4.4% |
| Iter. basic | Adv. training | top 1 | 77.4% | 30.0% | 25.2% | 23.5% | 23.2% |
|  |  | top 5 | 93.9% | 44.3% | 33.6% | 28.4% | 26.8% |
|  | Baseline | top 1 | 78.3% | 31.4% | 28.1% | 26.4% | 25.9% |
|  |  | top 5 | 94.1% | 43.1% | 34.8% | 30.2% | 28.8% |

|  | source model | FGSM target model | | | | basic iter. target model | | | | iter l.l. target model | | | |
|---|---|---|---|---|---|---|---|---|---|---|---|---|---|
|  |  | A | B | C | D | A | B | C | D | A | B | C | D |
| top 1 | A (v3) | 100 | 56 | 58 | 47 | 100 | 46 | 45 | 33 | 100 | 13 | 13 | 9 |
|  | B (v3) | 58 | 100 | 59 | 51 | 41 | 100 | 40 | 30 | 15 | 100 | 13 | 10 |
|  | C (v3 ELU) | 56 | 58 | 100 | 52 | 44 | 44 | 100 | 32 | 12 | 11 | 100 | 9 |
|  | D (v4) | 50 | 54 | 52 | 100 | 35 | 39 | 37 | 100 | 12 | 13 | 13 | 100 |
| top 5 | A (v3) | 100 | 50 | 50 | 36 | 100 | 15 | 17 | 11 | 100 | 8 | 7 | 5 |
|  | B (v3) | 51 | 100 | 50 | 37 | 16 | 100 | 14 | 10 | 7 | 100 | 5 | 4 |
|  | C (v3 ELU) | 44 | 45 | 100 | 37 | 16 | 18 | 100 | 13 | 6 | 6 | 100 | 4 |
|  | D (v4) | 42 | 38 | 46 | 100 | 11 | 15 | 15 | 100 | 6 | 6 | 6 | 100 |

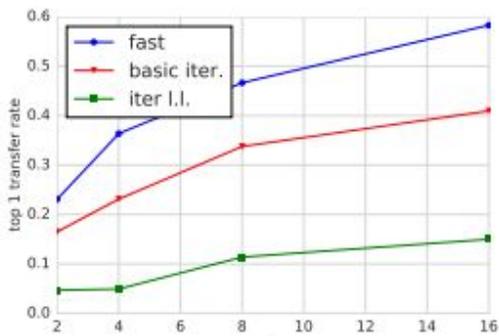

Top 1 transferability.

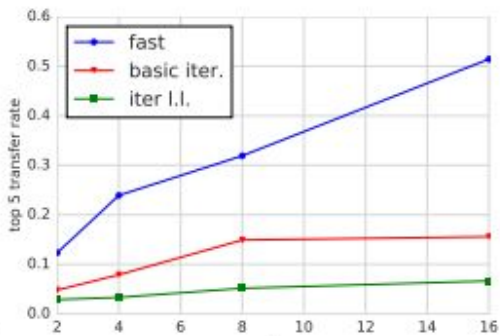

Top 5 transferability.

## 9 CONCLUSION

- It has been found empirically that adversarial examples can impact a deep learning convolutional network in misclassification of input images.
- Adversarial examples attack are direct attack on deep learning systems on the application layer. Mitigation on different layers of the network stack must be adhered to particularly on networked systems
- Adversarial examples are transferable from one model to another
- Adversarial training is considered as the most effective mitigation
- Standardized code hardening guidelines must be implemented in a global scale to reduce the risk of adversarial examples

**REFERENCES**

openai.com/blog/adversarial-example-research/

www.ibm.com/blogs/research/2018/05/clever-adversarial-attack/

arxiv.org/abs/1801.10578

arxiv.org/pdf/1801.10578.pdf

developer.ibm.com/articles/iot-lp101-connectivity-network-protocols/

github.com/tensorflow/cleverhans

arxiv.org/pdf/1705.07204.pdf

arxiv.org/pdf/1711.01991.pdf

www.ijcai.org/proceedings/2018/0543.pdf

arxiv.org/pdf/1712.07107.pdf

/arxiv.org/pdf/1611.01236.pdf

/medium.com/onfido-tech/adversarial-attacks-and-defences-for-convolutional-neural-networks-66915ece52e7